# Strong water absorption in the dayside emission spectrum of the planet HD 189733b

Carl J. Grillmair[1], Adam Burrows[2], David Charbonneau[3], Lee Armus[1], John Stauffer[1], Victoria Meadows[4], Jeffrey van Cleve[5], Kaspar von Braun[6], & Deborah Levine[1]

1. *Spitzer Science Center, 1200 East California Boulevard, Pasadena, California 91125, USA*

2. *Department of Astrophysical Sciences, Princeton University, Peyton Hall, Ivy Lane, Princeton, New Jersey 08544, USA*

3. *Harvard-Smithsonian Center for Astrophysics, 60 Garden Street, Cambridge, Massachusetts 02138, USA*

4. *Department of Astronomy, University of Washington, Box 351580, Seattle, Washington 98195, USA*

5. *Ball Aerospace & Technologies Corporation, PO Box 1062, Boulder, Colorado 80306, USA*

6. *Michelson Science Center, California Institute of Technology, Mail Stop 100-22, Pasadena, California 91125, USA*

**Recent observations of the extrasolar planet HD 189733b did not reveal the presence of water in the emission spectrum of the planet[1]. Yet models of such 'Hot Jupiter' planets predict an abundance of atmospheric water vapour[2]. Validating and constraining these models is crucial for understanding the physics and chemistry of planetary atmospheres in extreme environments. Indications of the presence of water in the atmosphere of HD 189733b have recently been found in transmission spectra[3,10], where the planet's atmosphere selectively absorbs the**

**light of the parent star, and in broadband photometry[4]. Here we report on the detection of strong water absorption in a high signal-to-noise, mid-infrared emission spectrum of the planet itself. We find both a strong downturn in the flux ratio below 10 μm and discrete spectral features that are characteristic of strong absorption by water vapour. The differences between these and previous observations are significant and admit the possibility that predicted planetary-scale dynamical weather structures[5] might alter the emission spectrum over time. Models that match the observed spectrum and the broadband photometry suggest that heat redistribution from the dayside to the night side is weak. Reconciling this with the high night side temperature[6] will require a better understanding of atmospheric circulation or possible additional energy sources.**

The extrasolar giant planet HD 189733b (ref. 8) is the most easily observable of the known transiting[9] exoplanets and has recently been the subject of intense scrutiny by both ground-based and space-based observatories. Owing to its relative proximity (19 pc), its high temperature (1201 K (ref. 10)), and its large size relative to its parent star ($R_p/R_* = 0.155$ (ref. 10)), the light from the planet can be easily distinguished in the infrared as a change in flux when the planet is observed during and outside of secondary eclipse (when the planet is hidden from view by the parent star). This favourable orbital alignment has enabled remarkably accurate measurements of temperature variations with planetary longitude[7] as well as a preliminary infrared emission spectrum of the dayside atmosphere[1].

Broadband filter observations of HD 189733b during primary transit using the Infrared Array Camera (IRAC) on the Spitzer Space Telescope recently yielded a flux deficit at 3.6 μm interpreted as absorption by water vapour[4]. However, an independent analysis of the same data concluded that the uncertainties in the observations are too large to support that interpretation[11]. Spectrophotometric observations of HD 189733b during primary transit with the NICMOS camera on the Hubble Space Telescope showed indications of both water vapour and methane[3], whereas ACS observations at visible wavelengths indicate a smooth but featureless trend towards smaller radii at longer wavelengths[12], suggesting the presence of small particle hazes. Detailed modeling[13] of mid-infrared broadband measurements taken during secondary eclipse[5] suggest that both water and CO are present in the planet's emission spectrum.

We observed HD 189733b with the Spitzer Space Telescope's Infrared Spectrograph[14] (IRS) during 10 secondary eclipses between mid-June and mid-December of 2007. Each exposure series was begun three hours before mid-eclipse and ended three hours after mid-eclipse. These data extend from 5 to 14 μm and provide the first mid-infrared spectrum below 7.5 μm. This is important because the strongest spectral signatures of absorption by water vapour and other molecular species occur between 5 and 10 μm. Removal of the effects of drifts, detector latencies, and telescope pointing oscillations has been described elsewhere[1]. In essence, our analysis measures the planetary light by subtracting the spectra obtained during secondary eclipse from the spectra taken before and after eclipse. In practice, we fit light curves to the time series data in each wavelength bin to make use of all available data, including the ingress and egress portions of the eclipse. The uncertainties are dominated by the short duration of the planet's passage behind the star (~1.5 hours for HD 189733b) during which we can measure the spectrum of the parent star in isolation. The Spitzer Space Telescope is currently the only facility capable of carrying out these observations.

In Figure 1 we show the planet/star flux ratio as a function of wavelength. Also shown in Figure 1 are the 3.6, 4.5, 5.8, 8.0, 16.0, and 24.0 μm broadband flux ratio measurements made using all three Spitzer Space Telescope's detectors (IRAC, IRS, and MIPS)[5,15] as well as an upper limit[16] on the flux ratio at 2.2μm obtained with the Near-Infrared SPectrograph on the Keck II telescope (W. M. Keck Observatory). The shape of the spectrum agrees quite well with that of the broadband measurements, although the 8.0 μm point lies slightly above the spectrum. The light-curve analyses applied to the two datasets are very similar (in as much as we treat the spectrograph as 128 independent photometers) and we know of no detector peculiarities that could give rise to a systematic offset. On the other hand, modestly significant differences have been noted in separate IRAC observations of the planet's 8.0 μm emission[5], suggesting that the planet's global dayside emission may actually be changing with time.

Also shown in Figure 1 are three detailed model atmospheres, all incorporating equilibrium molecular abundances at solar elemental metallicity. The model spectra have been smoothed and employ three different assumed values for the heat redistribution parameter, $P_n$ = 0.1, 0.15, and 0.3, and two values of the opacity, $\kappa_e$ = 0.0, 0.035 cm$^2$ g$^{-1}$, of a possible upper-atmosphere absorber that could be responsible for creating a slightly hotter upper atmosphere due to anomalous absorption of the incident stellar light[17]. The 3.6-μm/4.5-μm flux ratio and the decreasing planet/star flux ratio at wavelengths less than ~10 μm are signatures of the presence of gaseous water in absorption in the atmosphere of HD 189733b, with the possible presence of carbon monoxide a contributor near the IRAC 4.5 μm point[18,19]. The most interesting feature in our spectrum is the bump near ~6-6.5 μm in both the models and the data. This is the predicted[17-19] peak between the P and R branches of the $\nu_2$ vibrational bending mode of water vapour. It appears as an emission feature because the absence of a Q branch for this mode allows additional flux between the P and R branch absorptions.

We have investigated the significance of the 6.2 μm bump and the 7 μm trough, and a Kolmogorov-Smirnov test shows that the hypothesis that the 5.2-8 μm region of the spectrum could have been drawn from a featureless, straight-line, planet-to-star flux ratio can be rejected at the 98% confidence level. To establish the significance level of the 6.2-μm bump in a model-independent fashion, we excluded the data points from 5.9 to 6.6 µm and fitted a straight line to the remaining data between 5.2 and 8 µm. Using a $\chi^2$ test we find that we can reject the hypothesis that the observed data points between 5.8 and 6.6 µm are consistent with a straight line at the 99.992% probability level. A significant component of this result is contributed by an apparent peak at 5.9 µm, which is not due to water vapour and is not predicted in any of the model spectra. Goodness-of-fit statistics show no indication of poor light curve fits at or near this wavelength and we have no a priori reason to discount the data. However, if we exclude the data between 5.8 and 6.0 µm, we can still reject the null hypothesis at the 99.35% level.

While the models without an extra absorber (that is, with $\kappa_e = 0$) reproduce all the basic features of the observations, the absolute flux levels are much better reproduced by a model with an extra upper atmosphere absorber ($\kappa_e = 0.035$ cm$^2$ g$^{-1}$)[17]. This model also better reproduces the shallower spectral slope between 7 and 10 microns. However, such a low value of $\kappa_e$ does not produce a thermal inversion, merely a slightly hotter upper atmosphere. Unlike for HD 209458b (refs. 20, 21), the ratio between the IRAC 3.6 and 4.5 µm points is greater than one and the IRAC 5.8 µm point is not far above a straight line between 3.6 and 4.5µm. The former is consistent with the low (but non-zero) value of $\kappa_e$ needed for the best fit or with a greater CO abundance, while the latter might imply lower water abundance. Indeed, differences in the abundances, in the character and depth of heat redistribution, and in the nature and strength of a possible extra stratospheric absorber might be implicated in the emerging differences seen in the Spitzer data at secondary eclipse for the family of close-by extrasolar giant planets. Our models use a default prescription for the depth of heat circulation by super-rotational winds[17], so we suspect that a more comprehensive study that explores this dependence will further constrain the character of zonal heat transport. Moreover, performing 2D spectral calculations that better incorporate the integrated slant-angle effects on the planet's dayside hemisphere might also be expected to improve the agreement.

Our spectrum is best fit by models employing relatively low heat redistribution efficiencies. The apparent conflict between this and the high night side temperature of HD 189733b (ref. 7) and models that fit the broadband photometry has already been noted[5]. Our spectrum reinforces this conflict and, to this extent, supports suggestions[5] that the degree of heat redistribution may depend on atmospheric depth and could require three-dimensional modelling to be understood[22], or that HD 189733b has an internal energy source and radiates more energy than it receives through insolation.

Whereas the current measurements show a pronounced downturn below 10 μm, our previous IRS observations[1] of HD 189733b found the flux ratios to be essentially flat over the 7.5-14.5 μm wavelength range. Despite the smaller wavelength range, the downward slope at wavelengths less than ~10 μm should have been detectable in the earlier work. We note that a similar flat IRS spectrum was found for the giant exoplanet HD 209458b (refs. 23, 24), although this may have been due to the thermal inversion in its atmosphere[20,21]. A separate study[25] noted that a physically implausible source of opacity would have been required to reconcile the results of the earlier IRS observations with the 8μm photometric value[7], suggesting that the IRS systematic errors may have been larger than appreciated. Using the scatter in the present 7.5-14.5μm spectra as a measure of the uncertainty per transit observation, a comparison of our new result with the previous spectrum of HD 189733b yields a reduced $\chi^2$ of 1.9, indicating that the previous result is unlikely to have been due to a random departure from the mean. Similarly, if we fit slopes to the flux ratios as a function of wavelength, we find that the slope measured for the 2006 data departs at the 3σ level from the mean slope for the current observations. Whereas the average flux ratios among the current spectra differ at the 10% level, and the slopes in the flux ratios differ at the 30% level, our previous result is quite clearly out of character. The only difference between the present and the earlier data was in the use of longer exposure times and access to a wider wavelength range; the analysis procedures are essentially identical and a reanalysis of the 2006 data continues to produce a flat spectrum. There are currently no known systematic differences in the relative calibration of the shorter and longer exposures taken with the IRS.

A third possibility is that the upper atmosphere of HD 189733b may actually change with time. For tidally locked, slowly rotating hot Jupiters, models predict that dynamical weather structures will have planet-spanning scales[6]. Further observations over a longer time period will be required to substantiate such a hypothesis, and several more IRS transit observations of HD 189733b are planned for 2008.

**Acknowledgements** This work is based on observations made with the Spitzer Space Telescope, which is operated by the Jet Propulsion Laboratory, California Institute of Technology under a contract with NASA. Support for this work was provided by NASA through an award issued by JPL/Caltech. This study was supported in part by NASA grant NNGO4GL22G.

**Author Information** Reprints and permissions information available at www.nature.com/reprints. The authors declare no competing interests. Correspondence and requests for materials should be addressed to C.J.G. (carl@ipac.caltech.edu).

(Fig. 1) Comparison of spectral observations with broadband photometry and theoretical models of the dayside atmosphere of HD 189733b. The black points show the mean flux ratios for six $2^{nd}$-order spectra (5-8 $\mu$m) and four $1^{st}$-order spectra (7.5-14 $\mu$m). The data have been binned by a factor of four after light curve fitting (corresponding to two IRS resolution elements), and the plotted uncertainties reflect the standard error in the mean in each wavelength bin. The filled red circles show broadband measurements from ref. 5 at 3.6, 4.5, 5.8, 8.0, 16, and 24 $\mu$m (error bars on this data, s.e.). The upper limit at 2.2 $\mu$m derives from Keck spectroscopy[16]. The red, blue and green traces are atmospheric model predictions for three values of a dayside-nightside heat redistribution parameter and two values for the extra upper-atmosphere opacity, $\kappa_e$. The models predictions have not been scaled in any way.

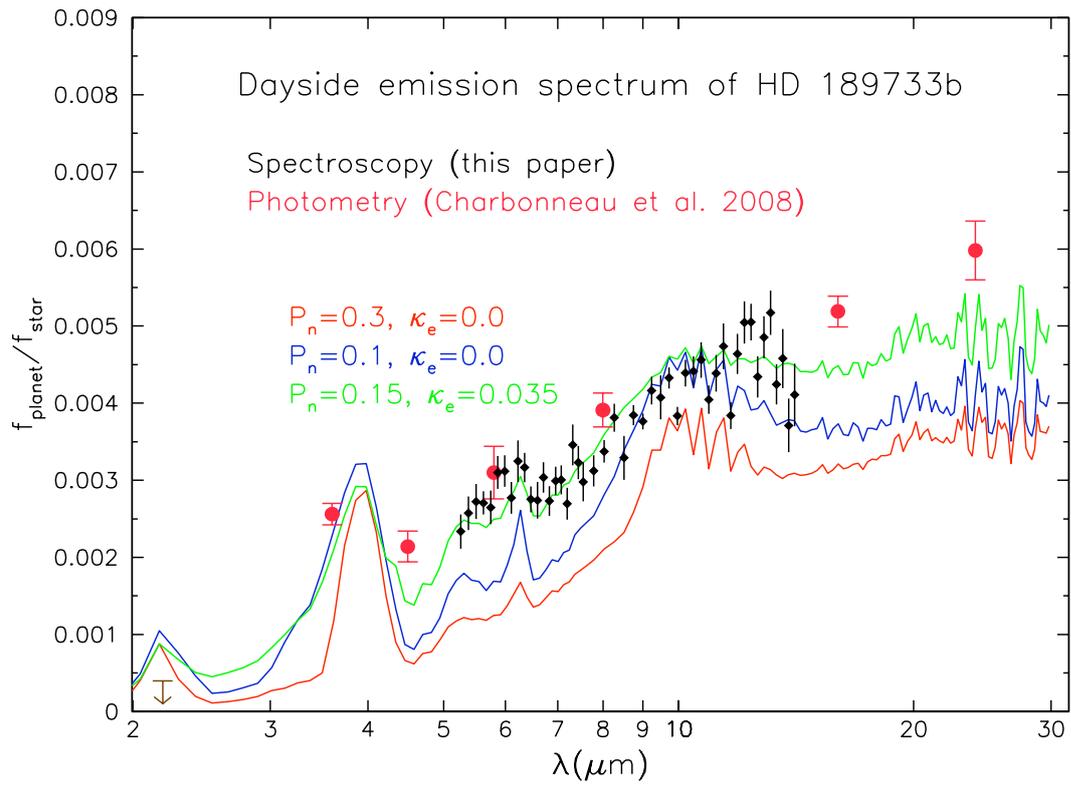